\documentclass[a4paper,11pt]{article}
\usepackage{pos}

\usepackage{amsfonts}
\usepackage{multirow}
\usepackage{makecell}
\usepackage{mathrsfs}
\usepackage{graphicx}
\usepackage{amsmath}
\usepackage{bm}
\usepackage{bbm}
\usepackage{color}
\usepackage{ulem}
\usepackage{array}
\usepackage{diagbox}
\usepackage{float}
\allowdisplaybreaks[4] 
\usepackage{setspace}
\def\OMIT#1{}

\def\hlinew#1{%
  \noalign{\ifnum0=`}\fi\hrule \@height #1 \futurelet
   \reserved@a\@xhline}
\usepackage{array}
\newcommand{\PreserveBackslash}[1]{\let\temp=\\#1\let\\=\temp}
\newcolumntype{C}[1]{>{\PreserveBackslash\centering}p{#1}}
\newcolumntype{R}[1]{>{\PreserveBackslash\raggedleft}p{#1}}
\newcolumntype{L}[1]{>{\PreserveBackslash\raggedright}p{#1}}

\newcommand{\beq}{\begin{equation}}
\newcommand{\eeq}{\end{equation}}
\newcommand{\bqa}{\begin{eqnarray}}
\newcommand{\eqa}{\end{eqnarray}}

\newcommand\fverb{\setbox\fverbbox=\hbox\bgroup\verb}
\newcommand\fverbdo{\egroup\medskip\noindent%
            \fbox{\unhbox\fverbbox}\ }
\newcommand\fverbit{\egroup\item[\fbox{\unhbox\fverbbox}]}
\newbox\fverbbox

\newcommand{\Rmnum}[1]{\expandafter\@slowromancap\romannumeral #1@}
\allowdisplaybreaks[2]

\title{Optimized QCD two-loop correction to 
exclusive double $J/\psi$ production at B factories}

\author*[a]{Wen-Long Sang}
\author[b]{Feng Feng}
\author[c,d]{Yu Jia}
\author[c,d]{Zhewen Mo}
\author[c,d]{Jichen Pan}
\author[c,d]{Jia-Yue Zhang}

\affiliation[a]{School of Physical Science and Technology, Southwest University,\\
  Chongqing 400700, China}

\affiliation[b]{China University of Mining and Technology,\\
Beijing 100083, China}

\affiliation[c]{Institute of High Energy Physics and Theoretical Physics Center for Science Facilities, Chinese Academy of Sciences,\\
 Beijing 100049, China}

\affiliation[d]{School of Physics, University of Chinese Academy of Sciences,\\
Beijing 100049, China}

\emailAdd{wlsang@swu.edu.cn}

\abstract{We report the calculation of the process $e^+ e^- \to J/\psi J/\psi$ up to next-to-next-to-leading order (NNLO) at a center-of-mass (CM) energy of $\sqrt{s}=10.58$ GeV. We employ an improved NRQCD factorization approach, decomposing the amplitude into photon-fragmentation and non-fragmentation components. The fragmentation contribution is determined using the measured $J/\psi$ decay constant, while the interference and non-fragmentation parts are computed at NNLO in $\alpha_s$ and lowest order in velocity. In this optimized scheme, both ${\cal O}(\alpha_s)$ and ${\cal O}(\alpha^2_s)$ corrections in the interference part are positive and exhibit good convergence. The non-fragmentation part is numerically insignificant. Our results indicate that with the projected 50 ${\rm ab}^{-1}$ dataset at \texttt{Belle 2}, the prospects for observing exclusive double $J/\psi$ production are very promising.}

\FullConference{The XVIth Quark Confinement and the Hadron Spectrum Conference (QCHSC24)\\
 19-24 August, 2024\\
 Cairns Convention Centre, Cairns, Queensland, Australia\\}

\begin{document}
\maketitle
\section{introduction}

In 2003, the \texttt{Belle} experiment searched for the double
$J/\psi$ production process $e^+ e^- \to  J/ \psi J/ \psi$ but found no clear signal~\cite{Belle:2004abn}. Instead, an upper limit was set $\sigma(e^+ e^- \to  J/ \psi J/ \psi )\mathcal{B}_{>2}< 9.1$ fb at the 90$\%$ confidence level, where $\mathcal{B}_{>2}$ denotes the branching fraction for final states
with more than two charged tracks.

Theoretical investigations of the $e^+ e^- \to  J/ \psi J/ \psi$ process have been conducted by various groups over the years. In 2002, Bodwin  et al. studied this process at the lowest order in the NRQCD approach~\cite{Bodwin:1994jh}, predicting a cross section of about
$8.7$ fb~\cite{Bodwin:2002fk}, which was later revised to $6.65$ fb~\cite{Bodwin:2002kk}. Davier et al.  considered photon fragmentation contributions and predicted a total cross section of about $2.38$ fb~\cite{Davier:2006fu}. Bodwin et al.  further included non-fragmentation contributions within the NRQCD framework, finding a destructive interference effect that reduced the cross section to about $1.69\pm 0.35$ fb~\cite{Bodwin:2006yd}.
Gong and Wang in 2008~\cite{Gong:2008ce}  computed the ${\cal O}(\alpha_s)$ corrections, finding that the NLO perturbative correction is negative and substantial, reducing the LO prediction from 
$7.4\sim 9.1$ fb to $-3.4\sim 2.3$ fb. Fan et al. later investigated the combined NLO perturbative and relativistic corrections~\cite{Fan:2012dy},  finding that the cross section can range from $-12$ fb to $-0.43$ fb in the fixed-order NRQCD approach, which is negative and sensitive to the charm quark mass and renormalization scale. However, by splitting the amplitude into photon-fragmentation and non-fragmentation parts, they obtained a positive cross section in the range of $1 \sim 1.5$ fb.

Given the wide range of predicted cross sections, precise theoretical predictions are crucial for guiding experimental searches. Motivated by the importance of ${\cal O}(\alpha_s)$ correction, we adopt an optimized NRQCD factorization approach to investigate the ${\cal O}(\alpha^2_s)$ correction
 to this process.
\section{Strategy of improved NRQCD factorization prediction for $e^+e^-\to J/\psi J/\psi$ }
\label{Fragmentation:production:rate}

\begin{figure}[h!]
\centering
\includegraphics[scale=0.7]{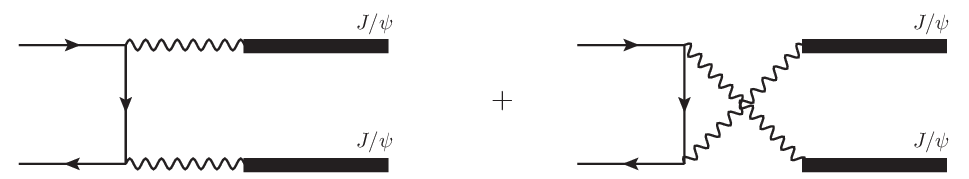}
\includegraphics[scale=0.7]{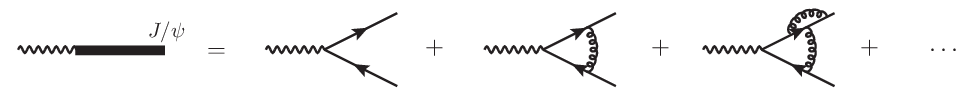}
\caption{Illustration of the $e^+e^-\to J/\psi+J/\psi$ process through two photon independent fragmentation. }
\label{diagrams:2Jpsi:from:photon:fragmentation}
\end{figure}

The double $J/\psi$ production in $e^+e^-$ annihilation at any order in $\alpha_s$ proceeds via either photon fragmentation, as illustrated in Fig.~\ref{diagrams:2Jpsi:from:photon:fragmentation}, or non-fragmentation channels, with the former dominating the latter. Following \cite{Bodwin:2006yd}, we decompose the production amplitude into fragmentation and non-fragmentation parts:
\beq
\frac{d\sigma}{d\cos\theta} = \frac{1}{2s}\frac{\beta}{16\pi} \frac{1}{4}\sum_{\rm spin} \left\vert {\cal M}_{\rm fr}+
{\cal M}_{\rm nfr} \right\vert^2.
\label{dsigma:dcos:decomposition}
\eeq
The fragmentation part is expressed in terms of $M_{J/\psi}$ and $f_{J/\psi}$~\cite{Bodwin:2006yd,Sang:2023liy}, while the non-fragmentation part is expressed using the charm quark mass and the NRQCD matrix element $\langle {\cal O}\rangle_{J/\psi}$. After squaring and summing over spins, the differential cross section is decomposed into fragmentation, interference, and non-fragmentation parts: 
\beq
\frac{d\sigma}{d\cos\theta} = \frac{1}{2s}\frac{\beta}{16\pi} \frac{e^8 e_c^4}{4}\bigg[\mathcal{C}_{\rm fr}f_{J/\psi}^4+\mathcal{C}_{\rm int} f_{J/\psi}^2 \frac{
\langle {\cal O} \rangle_{J/\psi}}{m_c}+\mathcal{C}_{\rm nfr}\bigg(\frac{\langle {\cal O} \rangle_{J/\psi}}{m_c}\bigg)^2\bigg],
\label{Optimized:NRQCD:factorization:formula}
\eeq
where the NRQCD matrix element is defined by
\beq
\langle {\mathcal O} \rangle_{J/\psi} \equiv
\vert \langle J/\psi(\lambda)|\psi^{\dagger}\boldsymbol{\sigma}\cdot\boldsymbol{\varepsilon}(\lambda)\chi|0\rangle\vert^2.
\eeq

The interference and non-fragmentation coefficients at lowest order in $v$ but through ${\alpha_s^2}$ are parameterized as:
\begin{subequations}
\bqa
\mathcal{C}_{\rm int} &=& {\cal C}_{\rm int}^{(0)}\bigg[1+\frac{\alpha_s}{\pi}\hat{c}_{\rm int}^{(1)}+\left(\frac{\alpha_s}{\pi}\right)^2
\bigg(\frac{\beta_0}{4}\ln\frac{\mu_R^2}{m_c^2}\hat{c}_{\rm int}^{(1)}+2\gamma_{J/\psi}\ln\frac{\mu_\Lambda^2}{m_c^2}+\hat{c}_{\rm int}^{(2)}\bigg)+\cdots\bigg],
\label{C:int:parametrization}
\\
\mathcal{C}_{\rm nfr}&=& {\cal C}_{\rm nfr}^{(0)}\bigg[1+\frac{\alpha_s}{\pi}\hat{c}_{\rm nfr}^{(1)}+\left(\frac{\alpha_s}{\pi}\right)^2
\bigg(\frac{\beta_0}{4}\ln\frac{\mu_R^2}{m_c^2}\hat{c}_{\rm nfr}^{(1)}+4\gamma_{J/\psi}\ln\frac{\mu_\Lambda^2}{m_c^2}+\hat{c}_{\rm nfr}^{(2)}\bigg)+\cdots \bigg],
\label{C:nfr:parametrization}
\eqa
\label{C:int:nfr:param}
\end{subequations}
where $\mu_R$ and $\mu_\Lambda$ are the renormalization and NRQCD factorization scales, respectively. $\beta_0=11C_A/3-2 n_f/3$ 
and $\gamma_{J/\psi}=-\frac{\pi^2}{12}C_{F}(2C_F+3C_A)$ is the two-loop anomalous dimension of the NRQCD vector current~\cite{Czarnecki:1997vz,Beneke:1997jm}.
Both $f_{J/\psi}$ and $\langle {\mathcal O} \rangle_{J/\psi}$ appear in the improved NRQCD factorization formula. However, these parameters are interrelated.  The decay constant $f_{J/\psi}$ encapsulates some perturbative effects and can be further factorized using NRQCD:
\beq
f_{J/\psi} = \sqrt{2 \langle {\mathcal O} \rangle_{J/\psi} \over M_{J/\psi}}\left[1+ \mathfrak{f}^{(1)}\frac{\alpha_s}{\pi}+\left(\frac{\alpha_s}{\pi}\right)^2
\left(\mathfrak{f}^{(1)}\frac{\beta_0}{4}\ln \frac{\mu_R^2}{m_c^2}+\gamma_{J/\psi}\ln\frac{\mu_{\Lambda}^2}{m_c^2}+ \mathfrak{f}^{(2)}\right) +\cdots \right]
+{\cal O}(v^2),
\label{decay:constant:NRQCD:fac}
\eeq
with $\mathfrak{f}^{(1)}=- 2 C_F$, $\mathfrak{f}^{(2)}=-43.3288$~\cite{Czarnecki:1997vz,Beneke:1997jm}.
The ${\cal O}(\alpha_s^3)$ correction~\cite{Marquard:2014pea,Feng:2022vvk} and ${\cal O}(\alpha^i_s v^2)$ ($i=0,1$) corrections~\cite{Keung:1982jb,Luke:1997ys}
have also been available.

\section{Leading-order cross section}
\label{LO:Improved:NRQCD}

\begin{figure}[h!]
    \centering
    \includegraphics[scale=0.7]{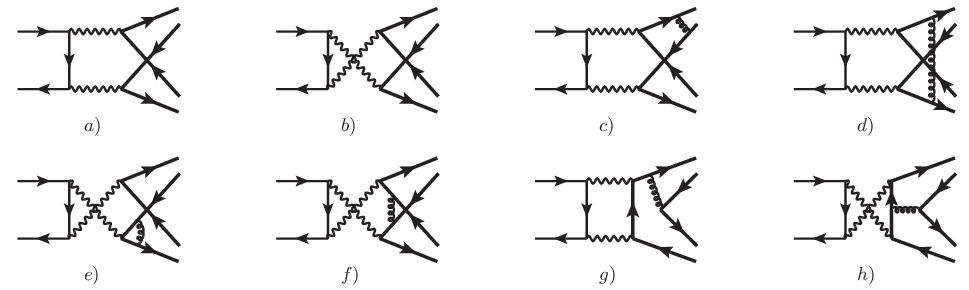}
    \caption{Non-fragmentation type of tree-level Feynman diagrams [$a)$ and $b)$], together with
    some sample one-loop non-fragmentation diagrams [$c)$ through $h)$]. }
    \label{Fig:nonfrag:diagrams:LO:NLO}
\end{figure}

The fragmentation diagrams are shown in Fig.~\ref{diagrams:2Jpsi:from:photon:fragmentation}, while the two tree-level non-fragmentation diagrams are depicted in Fig.~\ref{Fig:nonfrag:diagrams:LO:NLO}(a) and (b). The coefficient for the fragmentation term and the tree-level coefficients for the interference and non-fragmentation terms in Eq.~\eqref{Optimized:NRQCD:factorization:formula} can be found in Ref.~\cite{Sang:2023liy}.

By integrating Eq.~\eqref{Optimized:NRQCD:factorization:formula} over $\cos\theta$ from 0 to 1, we reproduce integrated cross sections:
\begin{subequations}
\begin{align}
&\sigma_\mathrm{fr} = \frac{32 \pi ^3 e_c^4 \alpha ^4 f_{J/\psi}^4 }{M_{J/\psi}^4 } {1\over s} \left[\frac{4+(1-\beta^2)^2}{1+\beta^2}
\ln \left(\frac{1+\beta}{1-\beta}\right)-2\beta \right],
\\
& \sigma_\mathrm{int} = -\frac{16 \pi ^3  e_c^4 \alpha ^4 f_{J/\psi}^2 \langle {\cal O} \rangle_{J/\psi}}{3 m_c^3 s^2} \left[
(5-\beta^2)(1-\beta^2)^2\ln \left(\frac{1+\beta}{1-\beta}\right)
+22 \beta -{40\over 3}\beta^3 + 2 \beta^5\right],
\\
& \sigma_\mathrm{nfr} = \frac{2048 \pi ^3 \alpha ^4 e_c^4 | \langle {\cal O} \rangle _{J/\psi}|^2}{45m_c^2 s^3}
\beta \left( 10- {20\over 3}\beta^2+\beta^4\right).
\end{align}
\label{total:X:section:tree:int:nfr}
\end{subequations}
Here the $J/\psi$ velocity $\beta$ is evaluated by replacing $M_{J/\psi}$ with $2 m_c$.
In contrast with the fragmentation part that asymptotically scales as $1/s$,
the interference part of the cross section exhibits a $1/s^2$ asymptotic decrease,  while the non-fragmentation part
exhibits a $1/s^3$ scaling.

\section{Higher-order radiative corrections}
\label{Higher:order:prediction:improved:NRQCD}

\begin{figure}[h!]
    \centering
    \includegraphics[scale=0.8]{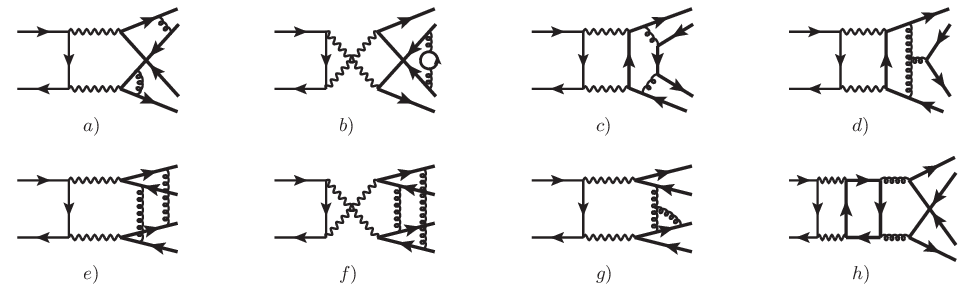}
    \caption{Some representative two-loop diagrams of non-fragmentation origin for $e^+e^-\to J/\psi+J/\psi$.}
    \label{Fig:nonfrag:diagrams:NNLO}
\end{figure}

To avoid double counting, we focus only on non-fragmentation loop diagrams. We start with the quark-level amplitude for 
$e^+e^-\to \gamma^*\gamma^*\to c\bar{c}({}^3S_1^{(1)})+c\bar{c}({}^3S_1^{(1)})$.
At lowest order in $v$, we neglect the relative momentum within each $c\bar{c}$ pair before performing the loop integration, effectively extracting the NRQCD SDCs directly from the hard loop region~\cite{Beneke:1997zp}.
We use dimensional regularization to handle both UV and IR divergences.
We generate approximately 24 one-loop and 506 non-vanishing two-loop non-fragmentation diagrams and their corresponding amplitudes using {\tt QGraf}/{\tt FeynArts}~\cite{Nogueira:1991ex,Hahn:2000kx}.
Representative one-loop and two-loop diagrams are shown in Fig.~\ref{Fig:nonfrag:diagrams:LO:NLO} and Fig.~\ref{Fig:nonfrag:diagrams:NNLO}. For simplicity, we omit "light-by-light" diagrams (e.g., Fig.~\ref{Fig:nonfrag:diagrams:NNLO}$h$)), which typically contribute negligibly to higher-order corrections in quarkonium production and decay~\cite{Feng:2015uha,Sang:2015uxg,Feng:2017hlu,Sang:2020fql,Yang:2020pyh,Feng:2022vvk}.

We use the covariant projector technique to ensure each $c\bar{c}$ pair carries the intended $^3S_1^{(1)}$ quantum number.
The trace over Dirac and $SU(N_c)$ color matrices is performed using the {\tt FeynCalc}/{\tt FormLink} packages~\cite{Mertig:1990an,Feng:2012tk}.
After applying integration-by-parts (IBP) reduction with {\tt Apart}~\cite{Feng:2012iq} and {\tt FIRE}~\cite{Smirnov:2014hma}, we obtain approximately 2400 two-loop master integrals (MIs). These MIs are computed using the powerful {\tt AMFlow} package~\cite{Liu:2017jxz, Liu:2021wks, Liu:2022mfb, Liu:2022chg}, which provides high numerical accuracy.

We perform field-strength and mass renormalization using the two-loop expressions for $Z_2$ and $Z_m$ from Ref.~\cite{Broadhurst:1991fy}, and renormalize the strong coupling constant under the $\overline{\rm MS}$ scheme to one-loop order. This eliminates the UV divergences in the two-loop SDCs. However, the renormalized two-loop corrections to $\mathcal{C}_{\rm int}$ and $\mathcal{C}_{\rm nfr}$
still contain uncancelled single IR poles proportional to ${\cal C}_{\rm int}^{(0)} \gamma_{J/\psi}$ and
to $2 {\cal C}_{\rm nfr}^{(0)} \gamma_{J/\psi}$, respectively.
This pattern is consistent with NRQCD factorization for double $J/\psi$ production at ${\cal O}(\alpha^2_s)$, as shown in Eq.~\eqref{C:int:nfr:param}.
The $\gamma_{J/\psi}\ln \mu_{\Lambda}^2$ terms in \eqref{C:int:nfr:param} cancel the $\mu_\Lambda$ dependence of the NRQCD matrix element, ensuring that the predicted cross section is independent of $\mu_{\Lambda}$.
Finally, we identify the non-logarithmic pieces in the two-loop SDCs, 
$\hat{c}_{\rm int}^{(2)}$ and $\hat{c}_{\rm nfr}^{(2)}$.

\section{Phenomenological results and discussions}
\label{}

\begin{figure}[h!]
\centering
\includegraphics[scale=0.8]{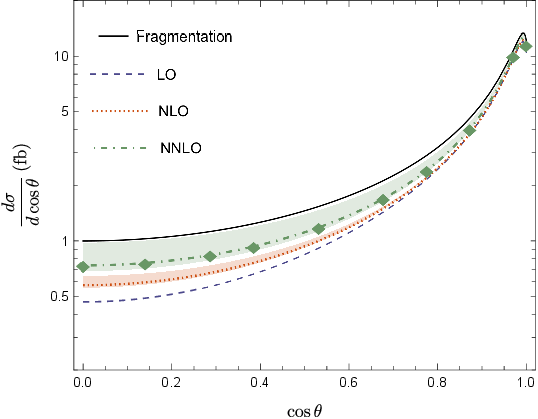}
\caption{Differential cross sections for $e^+ e^- \to J/\psi J/\psi$ against $\cos \theta$ at various perturbative accuracy
from improved NRQCD approach. We have fixed $\mu_{\Lambda} = 1\:\text{GeV}$, and taken the central value of $\mu_{R}$ to be $\sqrt{s}/2$.
The error bands of the NLO and NNLO predictions are estimated by sliding $\mu_R$ from $m_c$ to $\sqrt{s}$. }
\label{Plot:anuglar:distribution:double:Jpsi:Optimized:NRQCD}
\end{figure}

In our numerical analysis, we set  $\sqrt{s}=10.58$ GeV, $M_{J/\psi} = 3.0969\:\text{ GeV}$, $f_{J/\psi} = 403\:\text{MeV}$,
$m_c=1.5$ GeV, and $\langle {\cal O}\rangle_{J/\psi}(\mu_\Lambda=1\:{\rm GeV})=0.387\:{\rm GeV^3}$.
The default scale $\mu_R$ is chosen as $\sqrt{s}/2$, and we vary $\mu_R$ from $m_c$ to $\sqrt{s}$ to estimate the theoretical uncertainty in computing NLO and NNLO corrections. The running QCD coupling is computed to two-loop accuracy using the \texttt{RunDec} package~\cite{Chetyrkin:2000yt}.

In Fig.~\ref{Plot:anuglar:distribution:double:Jpsi:Optimized:NRQCD}, we show the angular distribution of $J/\psi$
at various perturbative orders using the improved NRQCD factorization. The LO prediction is significantly smaller than the fragmentation cross section due to destructive interference between the tree-level non-fragmentation and fragmentation amplitudes. However, both ${\cal O}(\alpha_s)$ and ${\cal O}(\alpha^2_s)$ corrections are positive in the improved NRQCD approach, showing good convergence. This contrasts sharply with the large negative NLO correction in the traditional NRQCD approach~\cite{Gong:2008ce}.
The fragmentation contribution dominates when the outgoing $J/\psi$ is collinear with the electron beam, while the interference term becomes significant as $\theta$ increases. The non-fragmentation term remains small throughout the range of $\theta$.

\begin{table}
\setlength{\tabcolsep}{13pt}
\renewcommand{\arraystretch}{1.6}
\begin{tabular}{ccccc}
\hline
  $\sigma$ (fb) & Fragmentation & LO & NLO & NNLO
\\\hline\hline
  Optimized NRQCD &
  \multirow{2}{*}{$2.52$} & $1.85$ & $1.93^{+0.05}_{-0.01}$ & $2.13^{+0.30}_{-0.06}$
\\
  Traditional NRQCD &
  & $6.12$ & $1.56^{+0.73}_{-2.95}$ & $-2.38^{+1.27}_{-5.35}$ \\
\hline
\end{tabular}
\caption{Integrated cross section of $e^+e^-\to J/\psi J/\psi$ at various perturbative accuracy.
The uncertainties are estimated by varying $\mu_R$ from $m_c$ to $\sqrt{s}$. }
\label{Integrated:Cross:Sections:B:factory}
\end{table}

In Table~\ref{Integrated:Cross:Sections:B:factory}, we summarize our predictions for the integrated cross section at $\sqrt{s}=10.58$ GeV at various perturbative accuracies. The NNLO prediction from the improved NRQCD approach is $2.13^{+0.30}_{-0.06}$ fb.
Unlike the negative cross section predicted by the standard NRQCD approach, our prediction using the optimized NRQCD approach is robust and reliable~\footnote{A different treatment is detailed in Ref.~\cite{Huang:2023pmn}.}. It is interesting to note that, our NNLO prediction for $\sigma(e^+e^-\to \gamma^*\gamma^*\to J/\psi J/\psi)$
is actually greater than the NNLO predictions for $\sigma(e^+e^-\to \gamma^*\to J/\psi+\chi_{c1})= {0.87}^{+0.19}_{-0.29}$ fb and
$\sigma(e^+e^-\to \gamma^*\to J/\psi+\chi_{c2})=  {0.73}^{+0.17}_{-0.27}$ fb at $B$ factories~\cite{Sang:2022kub}.

The \texttt{Belle} and \texttt{Belle 2} experiments have accumulated approximately 1500 ${\rm fb}^{-1}$ of data, which corresponds to about $3105\sim3645$ exclusive double $J/\psi$ events. Considering the branching fraction ${\cal B}(J/\psi\to l^+l^-) = 12\%$, this translates to $45\sim52$ four-lepton events from double $J/\psi$ decays. Assuming a reconstruction efficiency of 40\%, we expect to reconstruct about $18\sim21$ signal events. Given the potentially large background, identifying the double signal in the current dataset may be challenging. However, with the planned 50 ${\rm ab}^{-1}$ integrated luminosity at \texttt{Belle 2}, the prospects for observing exclusive double $J/\psi$ production appear promising in the near future.

\section{Summary}
\label{Summary}

We adopt an improved NRQCD factorization approach to calculate the  ${\cal O}(\alpha^2_s)$ correction to the process $e^+ e^- \to J/\psi+J/\psi$, by splitting the amplitude into photon-fragmentation and non-fragmentation parts. The fragmentation contribution is predicted using the measured $J/\psi$ decay constant, while the interference and non-fragmentation parts are computed at NNLO in $\alpha_s$ and lowest order in velocity. In this optimized scheme, both ${\cal O}(\alpha_s)$ and ${\cal O}(\alpha^2_s)$ corrections in the interference part are positive and show good convergence. The non-fragmentation part is numerically insignificant.
Our most accurate prediction is $\sigma(e^+ e^- \to J/\psi+J/\psi)=2.13^{+0.30}_{-0.06}$ fb  at $\sqrt{s}=10.58$ GeV, which is more reliable than the traditional NRQCD approach.
With the projected 50 ${\rm ab}^{-1}$ dataset at \texttt{Belle 2}, the prospects for observing exclusive double $J/\psi$ production are very promising.

\begin{acknowledgments}
The work of W.-L. S. is supported by the NNSFC Grant No. 11975187.
The work of F.~F. is supported by the NNSFC Grant No. 12275353 and No. 11875318.
The work of Y.~J., Z.~M., J.~P. and J.-Y.~Z is supported in part by the NNSFC Grants No.~11925506, No.~12070131001 (CRC110 by DFG and NSFC).
\end{acknowledgments}

\providecommand{\href}[2]{#2}\begingroup\raggedright

\end{document}